# On secret sharing for graphs
*Draft version*

KAMIL KULESZA, ZBIGNIEW KOTULSKI
*Institute of Fundamental Technological Research, Polish Academy of Sciences*
*ul.Świętokrzyska 21, 00-049, Warsaw, Poland, e-mail: {kkulesza,zkotulsk}@ippt.gov.pl*

Abstract: In the paper we discuss how to share the secrets, that are graphs. So, far secret sharing schemes were designed to work with numbers. As the first step, we propose conditions for "graph to number" conversion methods. Hence, the existing schemes can be used, without weakening their properties. Next, we show how graph properties can be used to extend capabilities of secret sharing schemes. This leads to proposal of using such properties for number based secret sharing.

## 1. INTRODUCTION

Graphs find applications in every field of computer science. Graph theory provides many NP class problems (see [9]), so it is not surprising that they find applications in cryptography and data security. Problem of building graph based security protocols was outlined in [12]. In the following paper authors discuss more specific problem of secret sharing protocols for graphs.

Everybody knows situations, where permission to trigger certain action requires approval of several selected entities. Equally important is that any other set of entities cannot trigger the action. Secret sharing allows to split a secret into different pieces, called shares, which are given to the participants, such that only certain group (authorized set of participants) can recover the secret.

Secret sharing schemes (SSS) were independently invented by George Blakley [2] and Adi Shamir [20]. Many schemes have been presented since, for instance, Asmuth and Bloom [1], Brickell [4], Karin-Greene-Hellman (KGH) [7]. Once secret sharing was introduced, people started to develop extended capabilities. Some of examples are: detection of cheaters and secret consistency verification (e.g., [18], [19], [21]), multi-secret threshold schemes (e.g., [18]), pre-positioned secret sharing schemes (e.g., [18]). The other class of extended capabilities focuses on anonymity, randomness and automatization for secret sharing procedures (e.g. [11]).

All of the mentioned above schemes were designed to work on the secret that is a number. As long as secret sharing was only toy for mathematicians, it did not pose a problem. In real-life applications secrets that are different then numbers are often found. Many of them can be seen as having structure of the graphs, for instance, consider complex structures like bio-chemical molecules or VSLI design (e.g. [10]). Such objects contain much more information that is given by simple formula. Such additional information can be often understood as graph related problem, e.g. finding optimal VSLI design is equivalent to finding minimal vertex coloring for some graph, see [10].

In the field of secret sharing, graphs were applied while studying access structures, for instance by Blundo, De Santis, Stinson and Vaccaro [3]. Although significant results were obtained, the role of the graphs was limited to modeling the data structures. Graphs provide variety of interesting properties, many of them resulting in the problems of NP class (e.g., [9]). At least few of these seem to be handy in secret sharing, especially for extended capabilities. When sharing the graph related secret, one has to consider two situations (or combination of them):
− sharing the graph structure,
− sharing graph properties (e.g. coloring), independently of graph structure.

Our approach is to find way to share graphs secrets using already available methods. In addition the graph being the subject of the protocol provides some benefits. One instance comes in the field of extended capabilities, while we other is more general. In fact, graph based secret sharing protocols result in the situation where one abstract type simultaneously underlies and models the protocol. This, in turn, allows one to see interaction of the protocol parts in a new light.

We propose conditions that should be fulfilled, to convert graph into number in order to apply existing secret sharing schemes. Once conversion is done, the graph properties hold for resulting number. In the Section 2 we will establish conversion conditions. This is followed by the example of conversion method in the next section. The Section 4 is devoted to problems of sharing graph properties, like vertex coloring. In the following section, few of the graph properties, useful for extended capabilities, are listed. In the Section 6 we deal with opposite situation, conversion of numbers into graphs in order to make use of graph based extended capabilities. The paper is summarized with conclusions and information on further research.

## 2. PRELIMINARIES

Notation:
Let $G(V,E)$ be the graph, where $V$ is set of vertices and $E$ is set of edges, with $|E|$ edges and $|V|$ vertices, $v_i$ denotes $i$th vertex of the graph, $v_i \in V$. Graph $G(V,E)$ will be referred as $G$ to simplify the notation. $\chi(G)$ is chromatic number for the graph $G$.

Define:
$M(G) = m$ is conversion function that takes graph $G$ and returns number $m$,
$M'(m) = G$ is the inverse of conversion function $M(G)$;

$|m|$ is the length of the number $m$.

The ultimate goal is to state conditions for $M(G)$, such that, once $G$ is converted into $m$, existing secret sharing schemes can be used. In addition, secret sharing scheme should not be weaken, even if $M(G)$ is made public. We start the discussion from case of sharing graph structure.

In order to share the graph structure, it is required for conversion function $M(G)$ to have the following properties:
- be injective (one-to-one),
- $M(G)$ and $M'(m)$ should be easy to compute,
- preferably be surjective (onto) the secret space of the secret sharing scheme.

The purpose of the last requirement is to maintain information theoretical properties of the secret sharing scheme (e.g. perfectness, see [18], [19]). Discussion on relaxation of that requirement will be carried out in the Section 3.4.

Let $\Gamma(|V|)$ be the number of all possible graphs having $|V|$ vertices. In other words, $\Gamma(|V|)$ represents number of binary sequences such that every bit corresponds to the possible edge in the $|V|$ vertices graph (1 denotes presence of the edge, 0 otherwise). It is well known that $\Gamma(|V|) = 2^{\frac{|V|(|V|-1)}{2}}$. Information contained in such string is sufficient to determine a graph. It can be seen as the special case of Goedel's numbering, see [5]. In the proposed approach all $M(G)$ fall into this category.

The knowledge of the graph structure, does not automatically yield all properties of the graph. In fact finding some of them can be the problem of NP class, see [9], [17]. In order to share graph properties, similar conditions apply. In addition, it is required that the inverse of that function yields information concerning only that property. It should support "non-disclosure rule". It means not to provide more information on the graph, than resulting from the information on that property. As the result sharing graph properties can be separated from sharing the graph structure. This allows to consider sharing secret (graph property) in two modes:
- when graphs structure is not known;
- when graph structure is known.

The second case is interesting when shared property belongs to the hard problems (see [9], [17], [10]). In the opposite case it would be easy to derive from the graph structure, hence treating it as the secret and sharing would be rather useless. Even if the property belongs to the class of difficult problems (say NP-complete), the proper care should be taken, whether the particular instance is not easily computable.

In the section 4 we give an illustrative example for sharing vertex coloring of the graph. Discussion of "non-disclosure rule" for that case will be also provided.

## 3. EXAMPLE OF CONVERSION METHOD

At the beginning simple scheme that allows conversion between graphs and numbers is presented. The conversion will be performed in two steps. First graph will be converted into the matrix, next matrix will be converted into the number. While more sophisticated methods can be used (e.g. [9]), the one chosen is a handy illustrative example.

## 3.1 Graph description

Graph $G$ is described by the square adjacency matrix $\mathbf{A} = [a_{ij}], i, j = 1, 2, ..., m$. The elements of $\mathbf{A}$ satisfy:
for $i \neq j$, $a_{ij} = 1$ if $v_i v_j \in E$ (vertices $v_i$, $v_j$ are connected by an edge) and $a_{ij} = 0$, otherwise;
for $i = j$, $a_{ii} = \alpha$, where $\alpha \in Z_k$ is the number of color assigned to $v_i$. In $Z_k$, $k \geq \chi(G)$ denotes the number of colors that can be used to color vertices of $G$ (in other words, $k$ is the size of the color palette).
In case that the graph coloring is not considered, $k=1$, and all entries on $\mathbf{A}$'s main diagonal are zero.

**Remark 1**
An adjacency matrix $\mathbf{A}$ determines the graph $G$. The opposite is not true. By permuting vertices of $G$ a variety of adjacency matrices can be produced. Hence additional information has to be provided to enforce injective property of the mapping $M(G)$. One of the possibilities is to fix order of $G$'s vertices.

**Example 1**
Take the graph $G$ with 4 vertices, colored with 3 colors:

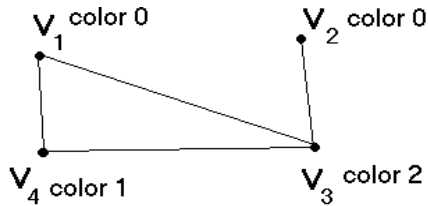

The adjacency matrix of the graph $G$ (only the graph structure, no colors) is presented on the left hand side, while the whole adjacency matrix $\mathbf{A}$ with encoded coloring is given on the right hand side.

|   | $v_1$ | $v_2$ | $v_3$ | $v_4$ |   | $v_1$ | $v_2$ | $v_3$ | $v_4$ |
|---|---|---|---|---|---|---|---|---|---|
| $v_1$ | 0 | 0 | 1 | 1 | $v_1$ | 0 | 0 | 1 | 1 |
| $v_2$ | 0 | 0 | 1 | 0 | $v_2$ | 0 | 0 | 1 | 0 |
| $v_3$ | 1 | 1 | 0 | 1 | $v_3$ | 1 | 1 | 2 | 1 |
| $v_4$ | 1 | 0 | 1 | 0 | $v_4$ | 1 | 0 | 1 | 1 |

∎

Coloring and the chromatic number are integral properties of any graph. Given the graph $G$, it is always possible to find its chromatic number and $n$-coloring.

## 3.2 Coding the matrix A

**A** is a symmetric matrix, hence having all the entries on the main diagonal and all the entries below main diagonal, one can describe whole matrix (and as the result graph *G*). Thus, it can be written as the sequence $a_{21}a_{31}a_{32}a_{41}a_{42}a_{43} \ldots a_{m(m-1)}a_{11}a_{22} \ldots a_{mm}$, where the first part ($a_{21}a_{31}a_{32}a_{41}a_{42}a_{43} \ldots a_{m(m-1)}$) corresponds to all the entries below main diagonal (graph structure), while the second ($a_{11} a_{22} \ldots a_{mm}$) to the main diagonal itself (the coloring).

**Example 1 (continuation)**
Coding matrix **A** we obtain

| $a_{21}$ | $a_{31}$ | $a_{32}$ | $a_{41}$ | $a_{42}$ | $a_{43}$ | $a_{11}$ | $a_{22}$ | $a_{33}$ | $a_{44}$ |
|---|---|---|---|---|---|---|---|---|---|
| 0 | 1 | 1 | 1 | 0 | 1 | 0 | 0 | 2 | 1 |

which yields *m=0111010021*

■

## 3.3 Coding the number as the graph

This can be done by converting number into binary form. Then the adjacency matrix **A** is encoded in the same way as depicted in the section 3.1. If the length *l* of binary number does not yield integer solution to the equation $l = \frac{(m-1)m}{2}$, then lacking matrix entries can be added.

## 3.4 Final remarks

Once graph is converted into a number traditional secret sharing schemes, as mentioned in the Section 1, can be applied. When this is done secret shares are standard "number-like" shares. In the time of the secret recovery they are pooled together using standard combiner algorithm. Resulting secret number is converted again into the graph as described in the Section 3.3.
We claim that presented conversion method comforts the requirements for $M(G)$ as stated in the Section 2:
− it is injective under condition that order of vertices is predefined;
− both $M(G)$ and $M'(m)$ are easy to compute
− is surjective when all graph configurations are permitted.
Once graph is converted into the number it can be shared using available secret sharing schemes. In order to maintain information theoretical properties of the secret sharing scheme (e.g. perfectness , see [18], [19]), one needs to carefully examine the space of possible *m* values. In many instances size of the space of possible *m* values will be smaller then cardinality of all numbers with $|m|$ digits. If this fact is not taken to account security of the secret sharing protocols can be significantly weakened.

**Example 2**
Consider set of connected graphs. As proposed in order to describe connected graph on $|V|$ vertices, one needs $m = \Gamma(|V|)$ bits. The secret space is clearly much smaller, because all strings representing non-connected graphs need to be excluded. Adversary trying to guess the secret will have task much easier then checking all possible numbers with length $|m|$ digits. ∎

The third requirement for $M(G)$ can be relaxed under two conditions:

a. knowing restrictions of the graph $G$, the space of possible $m$ values should be described. Properties (eg., perfectness) of underlying number based scheme should be derived and analyzed for that space.
b. when $M'(m)$ is computed, resulting graph $G$ should be checked whether it comforts imposed restriction.

If stated above conditions are properly applied resulting secret sharing scheme will not weaken properties of underlying number based scheme. There is also the other side of the coin, the restrictions allow for some added value. In the Section 5 we outline how to use them towards building extended capabilities into the secret sharing scheme.

## 4. SHARING GRAPH PROPERTIES

We use graph vertex coloring to illustrate how to share graph properties. Further in the section we assume (unless stated otherwise), that for the given graph $G$ finding its coloring is hard to compute. By hard we mean lack of polynomial algorithm for the problem, see [17].

### 4.1 Vertex coloring for graphs with unknown structure

As described in the section 3.2 vertex coloring of the graph $G$ can be written as the vector $a_{11} a_{22} ... a_{mm}$ with entries from main diagonal of the matrix **A**. Again, taking into account that any vector is possible (when any graph configuration is acceptable and graph structure is not known), all secret sharing methods suitable to share number can be applied.

When underlying secret sharing scheme is perfect, non-disclosure rule is supported. This results from the definition of perfectness, that participants obtain no information on the secret from their shares, till the recovery of the secret.

In above construction the only information available, when the secret is recovered, is the coloring of the graph $G$.

It should be emphasized that, in a general case, one can share only partitioning graph's vertices into $n$ sets (proper $n$-coloring for the graph), where $n = \chi(G)$, not a particular color-to-vertex assignment. It is due to the fact, that any secret participant can modify his share by adding component-wise a constant to every digit in the number. In such a case:

− a particular color-to-vertex assignment will be modified,

- the partitioning graph's vertices into *n* sets (proper *n*-coloring for the graph) can remain valid.

As the result this attack works when the secret is any minimal vertex coloring of the graph. This is the case when secret is solution to some problem and multiple solutions are permitted. Tampering with the secret shares by the participants does not result in any damage as long as recovered secret is valid minimal coloring of the graph. An instance of such problem will be presented in the Section 5.

When there is need to share particular color to vertex assignment, then the method to verify shares consistency needs to be added. This problem can be address by mean of Verifiable Secret Sharing, see [18].

In some special cases custom solutions can be designed, as discussed next.

## 4.2 Vertex coloring for graphs with known structure

When graph structure is known similar method to one described above can be used. However one should note that information contained in the graph structure severely limits the secret space. The particular secret space needs to be individually examined. Although approach outlined in the preceding section works, there are also other options to be considered. An interesting approach is presented in [14]. In this case assumption, that for the given graph *G* finding its coloring is hard, is inverted. Actually method is based on the fact that in some cases graph's vertices can be partitioned into different types depending on coloring properties. In such case it is relatively easy to find proper minimal coloring, but finding particular color to vertex assignment still can be difficult. The idea is to separately share information on vertices belonging to the different types of coloring sets. Only proper combination of secret resulting from the different types of coloring sets allows to recover particular vertex coloring of the graph *G*. As the result:

- sharing of single secret is turned into the series of combinatorial problems, that are easy for quantitative analysis;
- information theoretical properties can be derived and proof of security can be stated;
- method provides opportunity to share multiple secrets within same scheme.

Since graph structure is assumed to be known hence discussion of non-disclosure rule boils down to careful analysis of information theoretical properties (i.e. how much information about particular coloring is released when some of the secrets are recovered).

Authors considered adoption of this method for the graphs with unknown structure. Unfortunately revealing information about partition of the graphs into types of coloring sets leaks information on the graph structure. Yet, it seems possible to design computationally secure scheme. This line of research was stopped, because on the ground of information theory such scheme will be always considered weaker, then perfect one. Nevertheless it cannot be ruled out that demand for such solution will appear, resulting from some other consideration (e.g. extended capabilities).

# 5. ON GRAPH RELATED PROBLEMS, THAT ARE USEFUL IN SECRET SHARING

Graph theory provides us with variety of interesting problems, many of which are known to be of NP class (e.g., [9], [17]). Few instance are: graph coloring, graph isomorphism and Hamiltonian path.
In this section we provide examples, how to use some of these problems in secret sharing, with special emphasize on extended capabilities. It is interesting to note that often secret sharing is simultaneously applied to both structure and graph related property. In such case non-disclosure principle is used to manage interaction between this two.

## 5.1 Graph coloring and Verifiable Secret Sharing (VSS)

Soon after secret sharing schemes emerged, it has been realized that they were vulnerable to misbehaving protocol parties. Verification was introduced to protect against cheating participants, for instance see [21]. Unfortunately, usually it comes at the price. This fact is related to the paradox stated by David Chaum, that no system can simultaneously provide privacy and integrity.
 At ESORICS2002 in Zurich a verification method that works for any underlying secret sharing scheme was described ([16]). It is based on the concept of verification sets of participants, related to an authorized set of participants. The participants interact (with no third party involvement) in order to check validity of their shares before they are pooled for secret recovery. Verification efficiency does not depend on the number of faulty participants. One of the pillars of the method is the use of a proper verification function; a very promising one results from the graph coloring check-digit scheme described in [13]. This proposal requires conversion of the given number into a graph and checking its vertex coloring on both sides of the communication channel. The quantitative argument presented shows that the feasibility of the proposed scheme increases with the size of the number whose digits are checked, as well as, overall probability of digits errors.
Joining both results ([16], [13]) produces a graph-based shares verification method, which was less formally described in [15]. The method depends heavily on graph coloring properties that in turn are handy in the formal security analysis. To some extent it seems even to bypass (or at least weaken) the Chaum paradox. In the case of the described method one does not get a free lunch, but at least can have a free starter.
In the limiting case, combining verification with graph coloring check-digit , approaches idea of zero-knowledge argument (e.g.,[18]). It has been shown that every NP problem can be converted in to zero-knowledge proof (see, [6]). These observations enable us to draw conclusion that other graph related problems, with already constructed zero-knowledge proofs (for instance graph isomorphism), can be adapted for the shares verification method.

## 5.2 Restrictions on $M(G)$ based Verifiable Secret Sharing

The idea is straightforward. Consider sharing graph $G$, as described in the Section 3.4. Further assume that some restrictions (e.g. connectivity) are imposed on $G$. Verification comes after secret $m$ is recovered. Prior to use its validity is tested using mapping $M'(m)$. If $M'(m)$ maps into permitted graph type, then recovered secret can be valid.

Probabilistic verification protocol of this type was described in [16].
Proposal stated in this paragraph can be extended to use restrictions on $G$ and related mappings as the verification function for approach outlined in [16].

## 5.3 Public-key cryptosystems

In [8] Koblitz describes public-key cryptosystem "Polly Cracker", that uses graph 3-coloring . For his implementation the public key is the graph structure, while private key is graph 3-coloring. Hence in order to share the private key one has to be able share graph coloring. This very much the case described in the Section 4.
In addition, the fact, that any proper 3-coloring of the given graph can be a private key, yields greater flexibility when comes to construction of authorized sets of participant and resulting access structure (see [19]).
Whole problem was described and analyzed in [14].

## 5.4. Multi-secret schemes

Separate sharing of graph structure and it's properties, when combined in one protocol, allows to build multi-secret scheme. It provides different authorized sets of participants with different secrets. For instance, one set can recover the structure of the graph, while the other assignment of colors to the vertices. Only, when both groups cooperate graph together with vertex coloring can be obtained. Once hierarchy of graph properties is established, it can be turned into multi-secret threshold schemes (e.g., [18]). In such scheme, the greater number of participants cooperate, the more information about the graph can be recovered. For instance, consider DNA molecule. There are at least 4 levels on which that structure can be described, starting from the number of particular nucleotides (1st level structure) up the way that whole helix is folded (4th level structure). One can design a 4-thresholds scheme that links the number of participants taking part in the protocol with the level of DNA structure that they can recover.

## 6. FROM NUMBERS TO GRAPHS

Till now we were concerned with sharing objects of graph type. In preceding sections it was demonstrated that such objects can introduce additional features in secret sharing.

This leads to the question whether existing secret sharing schemes can benefit from those features? To describe problem by mean of example, can we use VSS, described above, for secret shares consisting of ordinary numbers? In other words, can we go the other way around and share a number the same way as a graph? In general context, answer to this questions is positive, provided proper mapping from numbers to graphs is found.

It is nothing different from $M'(m)$ discussed in the Section 2. In practice both mappings $M'(m)$ and $M(m)$ are needed. This requirement comes from the expected mode of operation:
- number will be converted into a graph $G$ using $M'(m)$;
- some operations will be performed on $G$ in order to support additional features/extended capabilities;
- resulting graph (say $G'$) has to be converted into the number again.

Natural way on conversion is to use $M(m)$ as defined in the Section 2. Special care has to be take for injective property of the mapping $M(G)$, as discussed in the Remark 1. The sample list of the benefits can be found in [16].

# 7. CONCLUDING REMARKS AND FURTHER RESEARCH

In the paper it was demonstrated how to share the secret, that is, a graph. Conversion into a number allows the use any of existing secret sharing schemes. Information theoretical capabilities for the chosen scheme should be preserved, provided that properties of $M(G)$ hold. New possibilities resulting from graph properties can be also applied to secrets that are numbers.

Further research will concentrate on:
- optimizing the conversion methods,
- working on more applications of graph properties to secret sharing and their applications to other objects ,
- formulation of the results in the Information Theory language,
- searching for alternative methods to share secret that is a graph.

Last one is really challenging task, because will allow to set new paradigms in secret sharing. The most promising are: visual cryptography methods and applications of the random graphs.

# ACKNOWLEDGEMENT

First version of this paper was prepared during visit to Rhodes University, Grahamstown, South Africa.

# 8. REFERENCES

[1] Asmuth C. and Bloom J. 1983. 'A modular approach to key safeguarding'. IEEE Transactions on Information Theory IT-29, pp. 208-211.
[2] Blakley G.R. 1979. 'Safeguarding cryptographic keys'. Proceedings AFIPS 1979 National Computer Conference, pp. 313-317.
[3] Blundo C., Giorgio Gaggia A., Stinson D.R. 1997.'On the Dealer's randomness required in secret sharing schemes'. Designs, Codes and Cryptography 11, pp. 107-122.
[4] Brickell E.F. 1989. 'Some ideal secret sharing schemes' Journal of Combinatorial Mathematics and Combinatorial Computing 6, pp. 105-113.
[5] Goedel K. 1931. 'Uber formal unentscheidbare Satze der Principia Mathematica und verwander Systeme I'. Monatshefte fur Mathematik und Physik, 38, pp. 173-198.
[6] Goldreich O., Micali S., Wigderson A. 1991. 'Proofs that yield nothing but their validity or all languages in NP have zero-knowledge proof systems'. Journal of the ACM, 38(1), pp.691-729.
[7] Karnin E.D., J.W. Greene, and Hellman M.E. 1983. 'On secret sharing systems'. IEEE Transactions on Information Theory IT-29, pp. 35-41.
[8] Koblitz N. 1998. 'Algebraic Aspects of Cryptography', Springer-Verlag, Berlin.
[9] Korte B., Vygen J. 2000. 'Combinatorial Optimization, theory and algorithms'. Springer-Verlag, Berlin.
[10] Kubale M. (eds). 2002. 'Optymalizacja dyskretna. Modele i metody kolorowania grafów'. Wydawnictwa Naukowo-Techniczne, Warszawa
[11] Kulesza K., Kotulski Z. 2003. 'On Automatic Secret Generation and Sharing for Karin-Greene-Hellman Scheme', To appear in: J.Sołdek, L.Drobiazgiewicz, [ed.], Artificial Intelligence and Security in Computing Systems, Kluwer, pp. 281-292.
[12] Kulesza K., Kotulski Z. 2003.' Addressing new challenges by building security protocols around graphs'. To appear in Lecture Notes in Computer Science, post-proceedings of the 11th Cambridge International Workshop on Security Protocols, Sidney Sussex College , 2-4.04.2003.
[13] Kulesza K., Kotulski Z. 2002.'On graph coloring check-digit method'. Submitted for publication, available from the Los Alamos arXiv (http://arxiv.org/).
[14] Kulesza K., Kotulski Z. 2002. 'Secret sharing for n-colorable graphs with applications to public-key cryptography'. Available from the Los Alamos arXiv (http://arxiv.org/).
[15] Kulesza K., Kotulski Z. 2002. 'On the graph coloring check-digit scheme with applications to verifiable secret sharing'. Available from the Los Alamos arXiv (http://arxiv.org/).
[16] Kulesza K., Kotulski Z., Pieprzyk J. 2002. 'On alternative approach for verifiable secret sharing'. ESORICS2002, Zurich. Submitted for publication, available from IACR's Cryptology ePrint Archive (http://eprint.iacr.org/) report 2003/035.
[17] Molloy M., Reed B.2002.'Graph Colouring and the Probabilistic Method'. Springer Verlag, Berlin.
[18] Menezes A.J., van Oorschot P. and Vanstone S.C. 1997. 'Handbook of Applied Cryptography'. CRC Press, Boca Raton.
[19] Pieprzyk J., Hardjono T. and Seberry J. 2003. 'Fundamentals of Computer Security'. Springer-Verlag, Berlin.
[20] Shamir A. 1979. 'How to share a secret'. Communication of the ACM 22, pp. 612-613.
[21] Stadler M. 1997. 'Publicly verifiable secret sharing'. Lecture Notes in Computer Science, pp.190-199 (Advances in Cryptology – EUROCRYPT'96).